\documentclass{osa-article}

\journal{oe}

\articletype{Research Article}

\begin{document}

\title{Attosecond resolution from free running interferometric measurements}

\author{Constantin Kr\"uger\authormark{1}, Jaco Fuchs\authormark{1,*}, Laura Cattaneo and Ursula Keller}

\address{Physics Department, ETH Z\"urich, 8093 Z\"urich, Switzerland\\
\authormark{1} These authors contributed equally to this work\\}

\email{\authormark{*} jafuchs@phys.ethz.ch} 



\begin{abstract}
Attosecond measurements reveal new physical insights in photo ionization dynamics from atoms, molecules and condensed matter. However, on such time scales even small timing jitter can significantly reduce the time resolution in pump-probe measurements. Here, we propose a novel technique to retrieve attosecond delays from a well established attosecond interferometric technique, referred to as Reconstruction of Attosecond Beating By Interference of Two-photon Transition (RABBITT), which is unaffected by timing jitter and significantly improves the precision of state-of-the-art experiments. We refer to this new technique as the Timing-jitter Unaffected Rabbitt Time deLay Extraction method, in short TURTLE. Using this TURTLE technique we could measure the attosecond ionization time delay between Argon and Neon in full agreement with prior measurements. The TURTLE technique allows for attosecond time resolution without pump-probe time delay stabilization and without attosecond pulses because only a stable XUV frequency comb is required as a pump. This will more easily enable attosecond measurements at FELs for example and thus provide a valuable tool for attosecond science. Here we also make a MATLAB code available for the TURTLE fit with appropriate citation in return.
\end{abstract}


\section{Introduction}
The discovery of high harmonic generation (HHG) \cite{ferray1988} and the better understanding of phase matching \cite{Huillier1991} offered access to ultrabroad optical pulses in the extreme ultraviolet (XUV) range. Given that HHG is a coherent nonlinear process, attosecond pulse generation was predicted \cite{antoine1996}, however, the pulse duration measurement became the new challenge. This was resolved with the RABBITT technique (Reconstruction of Attosecond Beating By Interference of Two-photon Transitions) \cite{Paul2001,Muller2002}, which was used to measure the first attosecond pulses within an attosecond pulse train (APT) repeating every half infrared (IR) laser cycle used for the HHG. Single attosecond pulses (SAPs) then were first measured with the attosecond streak camera technique \cite{Itatani2002}. Both the RABBITT and the attosecond streaking techniques are based on a pump-probe scheme, where an XUV attosecond pump pulse initiates electron dynamics and an IR probe pulse interacts with the temporal evolution of the released electrons as the time delay between pump and probe is varied. The RABBITT technique uses an APT in combination with a weak ($<10^{11}$ W/cm$^2$) and typically long IR ($\approx30$ fs) pulse. For attosecond streaking a linear polarized few-cycle IR probe pulse is used together with a SAP. These two techniques not only have been used to characterize attosecond pulses but also to resolve electron ionization dynamics in atoms \cite{Schultze2010,Drescher2002,Goulielmakis2004,Dahlstrom2012,Klunder2011,Palatchi2014,Isinger2017,Cattaneo2016,Fuchs2020}, small molecules \cite{Vos2018,Haessler2010,Huppert2016}, up to biomolecules \cite{Calegari2014} and solids \cite{Kruger2011,Cavalieri2007}. Streaking traces have to be analysed in the time-domain and relay on reconstruction algorithms with different approximations \cite{Mairesse2005,Gagnon2008,White2019,Pedrelli2020}, whereas the interference nature of the RABBITT-method allows for a more direct retrieval of the spectral phases \cite{Cattaneo2016,Isinger2019}.

However, both methods require an attosecond pump-probe delay control, which remains challenging and any XUV-IR pulse timing jiiter with uncontrolled fluctuations of the pump-probe delay can reduce the temporal resolution. Here we show that for the RABBITT technique we do not need any pump-probe delay control and attosecond time resolution can be achieved with a novel retrieval method, which is unaffected by jitter. This method, which we call Timing-jitter Unaffected Rabbitt Time deLay Extraction, in short TURTLE, proves, that neither attosecond pulses nor attosecond delay precision are needed to retrieve attosecond delays. In detail, we show that the sidebands (SBs) forming a RABBITT trace can be represented through an ellipse parametrization, which removes the pump-probe delay dependence in their analysis. An analogous method has previously been proposed for gravimeter measurements \cite{Foster2002} and is here adapted for attosecond science.

\begin{figure}[hb]
	\centering
	\includegraphics[width=\linewidth]{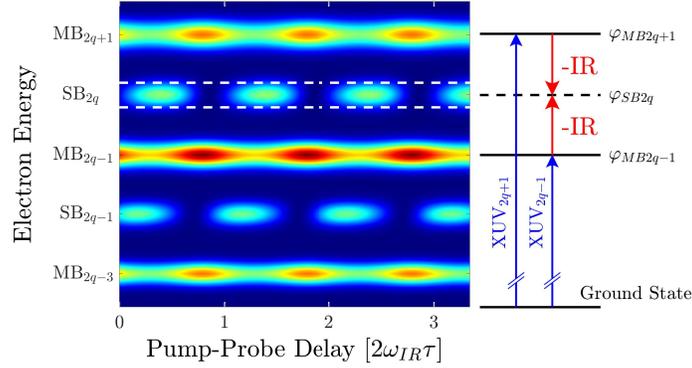}
	\caption{Example RABBITT spectrum, simulated within the strong-field approximation \cite{Lucchini2015}. The IR probe transfers electron population from the mainbands (MBs) to the sidebands (SBs), which oscillates with the delay between XUV pump and IR probe. The quantum pathways leading to the corresponding SB are illustrated schematically.} 
	\label{fig:Scheme}
\end{figure}

Figure \ref{fig:Scheme} shows an example RABBITT spectrum, where the photoelectron spectrum varies as a function of pump-probe delay $\tau$ between the XUV pump and the IR probe pulse. Upon absorption of an XUV photon, the photoemitted electron presents a kinetic energy which equals the difference between the XUV photon energy minus the ionization potential of the target. Hence, the photoelectron spectrum represents a replica of the XUV spectrum, shifted by the ionization potential and multiplied by the target specific ionization cross section. Due to the formation of an attosecond pulse train (APT) via high harmonic generation, the XUV spectrum consists of odd harmonics from the generating laser frequency, which results in discrete mainbands (MB) indexed 2q+1 ($q \in \mathbb{N}$) in the photoelectron spectrum. Due to the simultaneous presence of the IR probe, the absorption or stimulated emission of a further IR photon transfers electron population from the MBs to the SBs. These SBs oscillate as a function of the pump-probe delay as
\begin{equation}
    I_{2q}(\tau) \approx A_{2q}+B_{2q}\cdot \sin(2 \omega_{IR} \tau + \varphi_{2q}),
    \label{eq:SB_osc}
\end{equation}
where $A_{2q}$ and $B_{2q}$ are offset and amplitude of the oscillation and $\omega_{IR}$ corresponds to the IR frequency. The SB phase in turn contains different terms as follows:
\begin{equation}
    \varphi_{2q}= \Delta\varphi^{XUV}_{2q} + \Delta\varphi^{at}_{2q} + 2 \omega_{IR} \tau_0,
\end{equation}
where $\Delta\varphi^{XUV}_{2q}$ is the XUV contribution, corresponding to the phase difference between the XUV harmonics 2q-1 and 2q+1 (Fig. \ref{fig:Scheme}), $\Delta\varphi^{at}_{2q}$ is an atomic phase, which depends on the ionized target species, and the last term is an overall offset phase due to the unknown time zero $\tau_0$. In detail, the atomic phase corresponds to the phase difference due to the half-scattering process at the residual ionic potential between the two quantum paths leading to the SB formation \cite{Dahlstrom2012,Wigner1955,Pazourek2015}. Typically \cite{Klunder2011,Isinger2017,Cattaneo2016,Fuchs2020,Dahlstrom2012,Vos2018,Palatchi2014}, the SB phase is retrieved either via Fourier transform analysis or by performing a sinusoidal fit, which we further refer to as the sine-fit method. The retrieved phase can then be used twofold: (1) To estimate the attochirp, $\Delta(\Delta\varphi^{XUV})$, by using theoretical values for the atomic phase and comparing the phase difference of neighbouring SBs \cite{Muller2002,Dahlstrom2012}; or (2) to extract the atomic phase by comparing the phase difference between SBs stemming from different ionization channels or species but using the same SB order. Thus $\Delta\varphi^{XUV}$ cancels out and the difference of $\Delta\varphi^{at}$ can be isolated. This gives access to the photoionization time delays $\tau_{at}$ \cite{Klunder2011,Palatchi2014,Guenot2014,Isinger2017,Cattaneo2016,Fuchs2020,Vos2018} via
\begin{equation}
    \tau_{at} = \frac{\Delta\varphi^{at}_{SB1}-\Delta\varphi^{at}_{SB2}}{2 \omega_{IR}}.
    \label{eq:delay_phase_relation}
\end{equation}
In both cases, a relative phase between two SBs, \textit{i.e.} their phase difference $\Delta\varphi=\varphi_{SB1}-\varphi_{SB2}$, has to be determined such that the unknown time zero cancels out. Isinger et. al. \cite{Isinger2019} investigated the precision of the RABBITT technique and found that the precision of the retrieved phase is predominantly limited by the temporal stability of the pump-probe delay, \textit{i.e.} the timing jitter or time-drift between the two pulses. Although the active interferometric stabilization loops are used to reduce such effects, the pump-probe delay control is still the limiting factor for the precision in state-of-the-art experiments.

We will now present a new approach to overcome this issue, before comparing its precision with the sine-fit RABBITT extraction method in section 3. In section 4 we present a proof-of-principle experiment, where we measure RABBITT traces simultaneously in Argon and Neon gas targets \cite{Cattaneo2016,Guenot2014} without a delay scan and analyse the resulting SBs using TURTLE.

\section{The Method}
 
\begin{figure}[h!]
    \includegraphics[width=\linewidth]{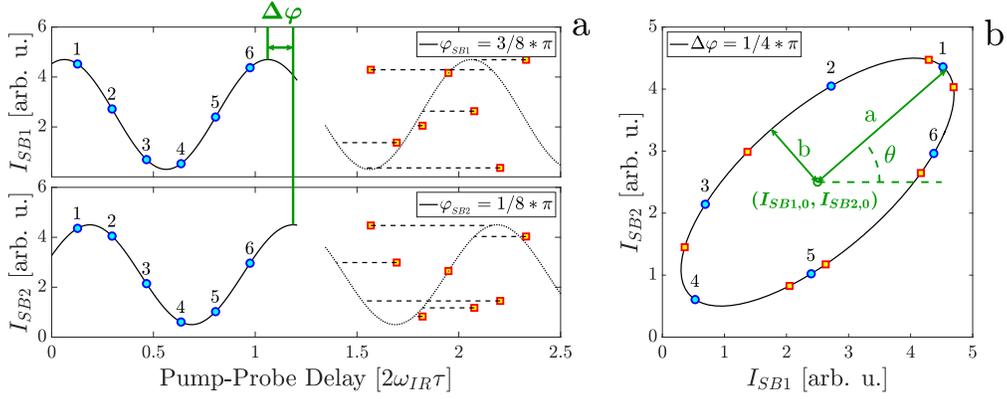}
    \caption{Illustration of the TURTLE method. (a) Two example SB oscillations with $\varphi_{SB1}=3/8$ and $\varphi_{SB2}=1/8$, where the blue circles correspond to jitter free acquisition and the red squares correspond to jitter affected acquisition. (b) The intensity correlation of SB 1 and 2 form the same ellipse for both acquisition types. The datapoints which are acquired jitter-free line up sequentially (counter clockwise) on the ellipse, indicated by the numbering.}
    \label{fig:Demo}
\end{figure}

To date attosecond time delays are calculated by fitting a sinusoidal function to the SB and extracting the calculated phase. Foster et. al \cite{Foster2002} proposed a method to extract the relative phase between two sinusoidal functions from coupled gravimeter interferometers, based on ellipse-specific fitting. This method is not affected by timing jitter. By analogy with SB intensities representing sinusoidal functions, any pair of SBs parametrizes an ellipse in the intensity correlation domain, \textit{i.e.} $I_{SB1}$ vs. $I_{SB2}$. Moreover, the relative phase $\Delta\varphi$ between the analysed SBs relates to the ellipse geometry as follows (Appendix A for the detailed derivation):
\begin{equation}
    \Delta\varphi= \varphi_{SB1}-\varphi_{SB2}=\arctan(\frac{b}{a} \cdot \tan(\theta))-\arctan(\frac{a}{b} \cdot \tan(\theta))+\frac{\pi}{2}
    \label{eq:delta_phi}
\end{equation}
where $a$ and $b$ are the major and minor half-axis, and $\theta$ corresponds to the tilt angle of the major half axis with respect to the x-axis, or respectively $I_{SB1}$-axis (Fig. \ref{fig:Demo}b).

Figure \ref{fig:Demo} illustrates the extraction of the relative phase between two SBs using the ellipse parametrization for two cases: the left part (Fig. \ref{fig:Demo}a) shows an equidistant sampling of two SBs, for which the phase difference can be retrieved by a sine-fit (blue samples enumerated from one to six). In the SB intensity correlation domain (Fig. \ref{fig:Demo}b) the data-points line up counter clockwise on the ellipse, from which $\Delta\varphi$ can be calculated using Eq. (\ref{eq:delta_phi}). The right part of Fig. \ref{fig:Demo}a shows data-points of the same SBs subjected to jitter (red squares). For these data-points, seemingly random distributed, the phase difference cannot be extracted using the sine-fit method. However, its intensity correlation remains unchanged due to the fact that both SBs are subjected to the same delay fluctuations. Accordingly, they satisfy the same ellipse equation as in the jitter-free case and, hence, $\Delta\varphi$ can be determined. Algebraically, this is given by the fact that in the ellipse parametrization $\Delta\varphi$ does not depend on the pump-probe delay $\tau$ anymore. Nevertheless, there are two constrains, which have to be satisfied. First, during the acquisition of a single data-point the jitter must be negligible. This is typically satisfied for shot-to-shot acquisition in large-scale facilities and for table-top experiments where the integration time per delay step is small compared to thermal drift. Second, $\Delta\varphi$ takes only values from $0$ to $\pi$, \textit{i.e.} it remains unknown which of the SBs is advanced. To overcome this ambiguity the rotation direction has to be captured (clockwise rotation means SB1 is delayed vs. SB2 and vice versa) which can be determined as long as the jitter is not larger than a full oscillation period. Since the formation of SBs only requires that the XUV bandwidth is at least twice the IR photon energy but does not depend on the pulse duration, neither attosecond pulses nor an attosecond delay control is required for measuring attosecond photoionization delays with TURTLE. Subsequently, we will compare the precision of TURTLE with the sine-fitting RABBITT technique before demonstrating its validity in a proof-of-principle experiment for unstabilized pump-probe delays. 

\section{Comparison between TURTLE and the sine-fit analysis}

\begin{figure}[h!]
    \centering
    \vspace{-0.3cm}
    \includegraphics[width=\linewidth]{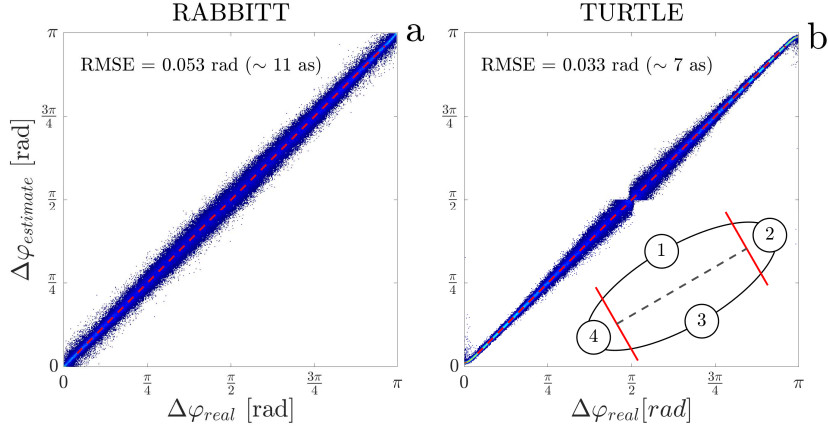}
  \caption{Comparison of the delay precision between the sine-fit method (a) and the TURTLE method (b) for 200'000 SB pair simulations. The phase shifts obtained by the two methods are plotted versus the input phase shift. TUTRLE provides a higher average precision (Root mean square error, RMSE of 0.033 rad) than the sine-fit method (RMSE of 0.053 rad).}
    \label{fig:simulation_comp}
\end{figure}
In order to compare the TURTLE method with the traditional sine-fit method, we consider two regimes: a high-jitter regime and a low-jitter regime.

In the high-jitter regime the pump-probe delay fluctuations are considered larger than half a laser cycle as it is the case for current state-of-the-art (seeded) free electron lasers (FELs). However, since shot-to-shot acquisition is possible in such facilities the TURTLE technique can enable the extraction of attosecond photoionization delays. In contrast, the traditional RABBITT sine-fit is not possible at all. The retrieved delay precision is only restricted by the flux stability and the acquisition time.

In the low-jitter regime, shot-to-shot pump-probe fluctuations are negligible and only small uncontrolled delay fluctuations (smaller than the tenth of a laser cycle) occur during the data acquisition at a fixed delay time. This is typically the case for table-top attosecond photoionization experiments. Accordingly, delay scans can be performed and both TURTLE and sine-fit can be used for the delay retrieval. Figure \ref{fig:simulation_comp} shows a comparison of the two-methods in a delay-scanned measurement under common experimental conditions. For this, we test both methods on sets of simulated SB pairs, generated via equation \ref{eq:SB_osc} with different intensities, amplitudes, and phases, the latter uniformly ranging from 0 to $\pi$. Each SB does 5 oscillations covering a pump-probe scan of 7 fs for an 800 nm laser wavelength with 50 steps of 130 as, in accordance with recent experiments \cite{Cattaneo2016,Isinger2017,Klunder2011,Palatchi2014,Fuchs2020}. Furthermore, we add 5\% of amplitude noise, and 80 as timing jitter to mimic experimental conditions. For the sine-fit extraction method, two sinusoidal functions are fitted simultaneously using a least-square minimization in order to guarantee a common oscillation frequency.
For the TURTLE method a novel ellipse fitting routine is developed which guarantees that successively acquired datapoints are fitted by the same section of the ellipse. Here we take advantage of the fact that in the delay-scanned acquisition neighbouring datapoints of the SB oscillation also correspond to neighbouring datapoints on the ellipse or, respectively, lie on the same side of the ellipse when there is a small amount of timing jitter (Fig. \ref{fig:Demo}). A detailed description of the fit implementation is given in Appendix B. As shown in Fig. \ref{fig:simulation_comp} TURTLE generally provides a higher precision in the phase retrieval than the sine-fitting method. In particular, using TURTLE the retrieved phases $\Delta\varphi$ for the set of 200'000 simulated SB pairs show a 63\% lower root mean square error (RMSE) compared to the sine-fit method. This can be explained by the fact that TURTLE is not affected by timing jitter. Indeed, when the same test is performed on simulations without jitter, both methods yield comparable precision.

Thus, TURTLE does not only enable the delay extraction in the high-jitter case, i.e. for FELs, but also provide a more accurate tool for table-top laser experiments which usually have lower timing jitter.

\section{Proof-of-principle Experiment}

\begin{figure}[h!]
    \centering
    \includegraphics[width=\linewidth]{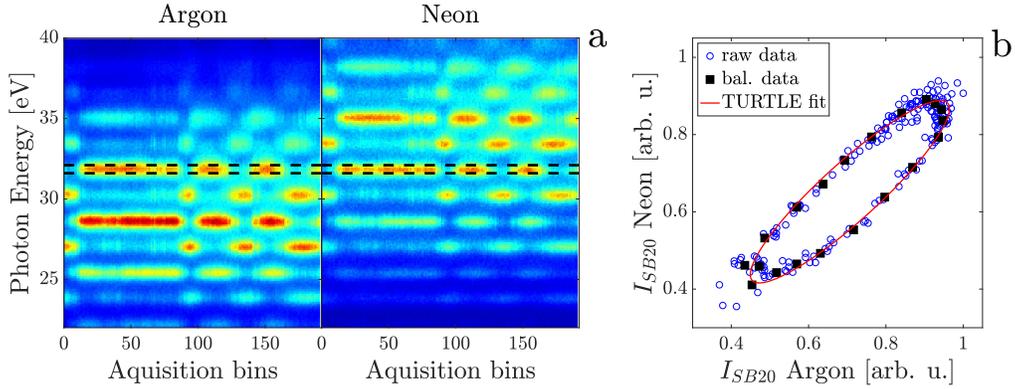}
    \caption{(a) Experimental RABBITT trace from Argon 3p (left) and Neon 2p (right) photoionization obtained by the thermal drift of the experimental setup. The black dotted lines indicate the energy range integrated to obtain the signal of SB 20. (b) Correlation of the SB intensities. The blue circles show the raw data, the black squares the balanced representation, and the red ellipse the TURTLE fit.}
    \label{fig:unstabilized_data}
\end{figure}
 
In a last step, we record a RABBITT trace using the same setup as described in \cite{Sabbar2014}, but without any stabilization of the pump-probe delay and using any delay scan. The thermal drift of the optical components forming the two arms of the pump-probe set-up are the main reason for an uncontrolled timing jitter between the two pulses. Figure \ref{fig:unstabilized_data} shows a 12-hours RABBITT trace measured in a gas mixture of Argon and Neon (1:1) without a delay scan. The coincidence detection of the cold target recoil ion momentum spectrometer \cite{Do2000} allows us to isolate RABBITT spectra for the two atomic species. The data are binned in time intervals of 4 min, which are short enough to resolve delay variations. Figure \ref{fig:unstabilized_data}b shows the intensity correlation of SB 20 for both atoms and the corresponding TURTLE fit for the phase retrieval. Since the drift is random, the obtained data-points do not sample the ellipse homogeneously as in a delay-scanned acquisition, thus a balancing of the raw data is required. The data points are averaged in 20 evenly spaced angular segments around the center of a primarily fitted ellipse and an average value is assigned for each angular segment. The final ellipse is then fitted on the balanced dataset. Although the balancing is not needed in our specific experiment, it might become more crucial in other measurements. We obtain a delay of 88 as $\pm$ 6 as at a photon-energy of 31.9 eV in excellent agreement with previous measurements (85 as $\pm$ 10 as, at 32.1 eV \cite{Cattaneo2016}). For the error analysis, we first calculate the standard deviation within each segment as well as its mean distance to the fitted ellipse to estimate the error of the fit. Consecutively, we perform an error propagation on Eq. (\ref{eq:delta_phi}) where we use the same error for the major and minor half axis, and neglect the error on the tilt angle $\theta$. This experiment demonstrates, that TURTLE is a viable method for retrieving attosecond time delays without the need for pump-probe delay scans with attosecond precision.

In particular, TURTLE enables the retrieval of attosecond delays in single-shot acquisition measurements with high XUV flux where shot-to-shot jitter can be as large as few femtoseconds, currently typically observed for FEL facilities \cite{Kang2019,Drescher2017,Danailov2014}. Note that a broadband XUV spectrum is still required in order to enable the interference of two-photon transitions. Therefore the TURTLE technique enables photoionization experiments beyond the typical RABBITT application for which the SB generating laser field is from the same source as the generating laser field for the HHG \cite{Schlaepfer2019}, which reduces their relative timing jitter in the measurements. 

\section{Conclusion}

We developed a novel analysis technique, TURTLE, which enables a jitter-free retrieval of attosecond photoionization delays from RABBITT measurements. Using a correlation representation of the SB intensities, the method not only allows for the retrieval of attosecond delays with higher precision than the sine-fit method but also opens up new experimental approaches. In particular, we demonstrate in a proof-of-principle experiment that TURTLE neither requires attosecond light pulses nor attosecond delay control in order to retrieve attosecond ionization delays. Solely by exploiting the unstabilized timing jitter in the experiment, we retrieve the Ar-Ne delay in excellent agreement with literature values. Our method not only supports state-of-the-art table top laser experiments but also paves the way for attosecond measurements at FEL facilities and thus provides a valuable tool to the field of attosecond science.

\section*{Appendix A: Derivation of equation \ref{eq:delta_phi}}
\label{Sec:Appendix_a}
\begin{figure}[!ht]
    \centering
    \includegraphics[width=\linewidth]{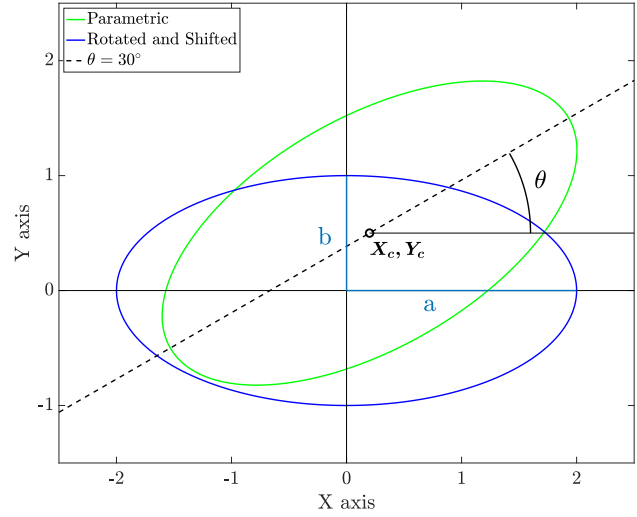}
    \caption{Illustration of the rotated and shifted ellipse.}
    \label{fig:Ellipse}
\end{figure}

The parametric equation of an ellipse in Cartesian coordinates with the major axis $a>0$ parallel to the x-axis and the minor axis $b>0$ parallel to the y axis is (Fig. \ref{fig:Ellipse}, blue ellipse):
\begin{align}
\begin{split}\label{eq:ellipse_easy}
x(t) &= x_c + a \cdot \cos(t)\\
y(t) &= y_c + b \cdot \sin(t),
\end{split}
\end{align}
where $t$ ranges from $0$ to $2\pi$. Applying a rotation by an angle $\theta$ and a shift by $x_c$ in x-direction and $y_c$ in y-direction the ellipse becomes (Fig. \ref{fig:Ellipse}, green ellipse)
\begin{align}
\begin{split}\label{eq:ellipse_tilted}
x(t) &= x_c + a \cdot \cos(\theta) \cdot \cos(t) - b \cdot \sin(\theta) \cdot \sin(t)\\
y(t) &= y_c + a \cdot \sin(\theta) \cdot \cos(t) + b \cdot \cos(\theta) \cdot \sin(t).
\end{split}
\end{align}
The ellipse rotation is illustrated in Fig. \ref{fig:Ellipse}. The rotation angle $\theta$ only ranges from $-\pi/2$ to $\pi/2$ in order to avoid ambiguities. Using trigonometric identities (linear combination of trigonometric functions) Eq. (\ref{eq:ellipse_tilted}) can be rearranged to
\begin{align}
\begin{split}\label{eq:ellipse_rearranged}
x(t) &= x_c + x_0 \cdot \sin[t + \arctan( \frac{b}{a} \cdot \tan(\theta))+\frac{\pi}{2}]\\
y(t) &= y_c + y_0 \cdot \sin[t + \arctan(\frac{a}{b}\cdot\tan(\theta))],
\end{split}
\end{align}
where $x_0=\sqrt{a^2 \cos^2(\theta)+b^2 \sin^2(\theta)}$ and $y_0=\sqrt{a^2 \sin^2(\theta)+b^2 \cos^2(\theta)}$. The ellipse parametric equation has the same structure as two SB signals (\ref{eq:SB_osc}):
\begin{align}
\begin{split}
I_{SB1}(\tau) &= A_{SB1} + B_{SB1}\cdot \sin(2\omega \tau + \varphi_{SB1})\\
I_{SB2}(\tau) &= A_{SB2} + B_{SB2}\cdot \sin(2\omega \tau + \varphi_{SB2}).
\end{split}
\end{align}
Hence, the phase difference $\Delta\varphi$ between two SBs can be identified as in Eq. (\ref{eq:delta_phi}).

\section*{Appendix B: TURTLE ellipse fit for delay-scanned acquisition}
\label{Sec:Appendix_b}

We separate the data-points in four segments along the major axis of the ellipse. Two edge segments (2 and 4) as well as an upper (1) and lower (3) center segment, illustrated by the inset in Fig. \ref{fig:simulation_comp}. Afterwards, we group consecutive data-points in batches, until the following point crosses a segment boundary (solid red line in the inset of Fig. \ref{fig:simulation_comp}). For example, five SB oscillations will result in $5\times4 = 20$ batches if there is no noise. Otherwise, there might be more batches, because amplitude fluctuations and jitter can cause multiple crossings. The data-points in the center batches (segments 1 and 3) are then fitted jointly either by the upper or the lower segment of the ellipse, depending on which minimizes the entire batch residual. This avoids that data-points in the center segments are not mistakenly fitted to the wrong side of the ellipse, which is especially important for noisy and narrow ellipses ($\Delta\varphi$ close to $0$ or $\pi$). The data-points in the edge batches (segments 2 and 4) are fitted individually by the closer ellipse part (upper or lower), because here it is not known a priory to which side of the ellipse they belong. To make the distinction between upper and lower part, the ellipse is rotated onto the x-axis, where both parts can be expressed by 
\begin{equation}
    \centering
    f_{up}(x) = -f_{down}(x) = b\cdot sin(arccos(\frac{x-x_0}{a}))
    \label{eq:fit_function}
\end{equation}
where $a$ and $b$ are the major and minor half-axis, and $x_0$ is the center of the rotated ellipse. The segment boundaries are given by $x=x_0\pm 0.5a$. Accordingly, the fit is conducted by minimising the least-square residual
\begin{align}
\begin{split}\label{eq:residual}
Res ={}& \sum_{i} min([-f(x_i)-y_i]^2,[f(x_i)-y_i]^2) \\
       & + \sum_{j} min(\sum_{k\in K_j} [-f(x_k)-y_k]^2,\sum_{k\in K_j} [f(x_k)-y_k]^2)
\end{split}
\end{align}
where $i$ runs over all data-points in the edge segments, $j$ runs over the center batches and $k$ runs over the $K_j$ data-points in each center batch. A MATLAB code for the TURTLE ellipse fit is provided online.

\section*{Funding}

Schweizerischer Nationalfonds zur F\"orderung der Wissenschaftlichen Forschung (NCCR MUST)

\section*{Disclosures}

The authors declare no conflicts of interest.


\bibliography{Phase_retrieval_paper}

\end{document}